\documentclass[proceedings, preprint]{rmaa}

\usepackage{paralist}
\usepackage{psfrag,color}
\usepackage{url}

\SetYear{2010}
\SetConfTitle{XIII Latin American Regional IAU}
\title{Correlations between radio emission of the parsec-scale jet and optical
nuclear emission of host AGN}

\author{
  J. Torrealba,\altaffilmark{1}
  T.~G. Arshakian,\altaffilmark{2}
  V. Chavushyan,\altaffilmark{1}
  and I. Cruz-Gonz\'alez\altaffilmark{3} }

\altaffiltext{1}{Instituto Nacional de Astrof\'isica \'Optica y Electr\'onica, Apartado Postal 51 y 216, 72000 Puebla, Pue, M\'exico (cjanet@inaoep.mx, vahram@inaoep.mx).}
\altaffiltext{2}{Max-Planck/Institut f\"ur Radioastronomie, Auf dem H\"ugel 69, 53121 Bonn, Germany (tigar@mpifr-bonn.mpg.de).}
\altaffiltext{3}{Instituto de Astronom\'ia, Universidad Nacional Aut\'onoma de M\'exico, Ciudad Universitaria, Apartado Postal 70--264,  04510, M\'exico D.F., M\'exico
(irene@astroscu.unam.mx).}

\shortauthor{Torrealba et al.}

\shorttitle{Radio--Optical correlations for compact AGN}

\listofauthors{J. Torrealba, T. G. Arshakian,  V. Chavushyan \& I. Cruz-Gonz\'alez}

\indexauthor{Torrealba, J.}
\indexauthor{Arshakian, T. G.}
\indexauthor{Chavushyan, V.}
\indexauthor{Cruz-Gonz\'alez, I.}

\resumen{Estudiamos las relaciones entre la emisi\'on  VLBA (Very Long Base Array) en radio a 15\,GHz y la emisi\'on nuclear \'optica a 5100 \AA\, para una muestra de 233 AGN dominados por el core  con jets relativistas. Para 181 cuasares, hay una correlaci\'on positiva significativa entre la luminosidad nuclear \'optica y las luminosidades VLBA totales del los n\'ucleos no resueltos  (en escalas de mili-arcosegundos) de los jets a 15\,GHz. La emisi\'on del continuo \'optico correlaciona con la emisi\'on del jet a 15\,GHz para 31 BL Lacs. Estas correlaciones confirman que la emisi\'on en radio y en \'optico est\'an amplificadas y se originan en la parte m\'as interna del jet a escalas de sub--parsecs en los cuasares, mientras que en los BL Lacs est\'as emisiones son generadas en el jet a escalas de parsecs. Estos resultados  est\'an de acuerdo con lo reportado previamente por \citet{arshakian10} para una muestra de 135 AGN.}

\abstract{We study the relation between the VLBA (Very Long Baseline Array) radio emission
at 15 GHz and the optical nuclear emission at 5100 \AA\, for a sample of 233 core-dominated AGN with relativistic jets. For 181 quasars, there is a significant positive correlation between optical nuclear emission and total radio (VLBA) emission of unresolved cores (on milliarcsecond scales) of the jet at 15 GHz. Optical continuum emission correlates with radio emission of the jet for 31 BL Lacs. These correlations confirm that the radio and optical emission are beamed and originate at sub-parsec scales in the innermost part of the  jet in quasars, while they are generated in the parsec-scale jet in BL Lacs. These results are in agreement with that reported earlier by \citet{arshakian10} for a sample of 135 AGN.}

\addkeyword{galaxies:active -- galaxies}
\addkeyword{galaxies: jets-- radio continuum:}
\addkeyword{galaxies -- quasars: general}

\begin{document}

\maketitle
\section{Introduction}
\label{sec:intro}
One approach to investigate the physical processes in active galactic nuclei (AGN) at scales not
reachable by present-day telescopes is to study the correlation between radiative energy in different frequencies. \citet{arshakian10} (hereafter A10) investigated the radio-optical correlation between the VLBA core emission at 15\,GHz and the optical nuclear emission at 5100\,\AA\ for a statistically complete sample of 135 compact jets to test a single production mechanism for radio and optical continuum emission on scales of submilliarcseconds. The present study is an extension of that research with a sample of 250 compact extragalactic radio sources at 15\,GHz compiled by \citet{kovalev05}, which includes 135 compact AGN from the flux-density limited MOJAVE-1 sample \citep{lister09}. The sample comprises 188 quasars, 36 BL Lacs, 20 radio galaxies, and 6 sources with no optical identification. 
The majority of AGN in the sample have relativistic jets oriented to the line of sight and have an unprecedented resolution of 0.5 milliarcseconds at 15 GHz. Note that this sample is not complete by the limiting radio flux.

\begin{table*}[!t]\centering
  \setlength{\tabnotewidth}{1.0\textwidth}
  \setlength{\tabcolsep}{1.1\tabcolsep}
  \tablecols{11}
  \caption{Kendall's $\tau$ correlation analysis between radio and optical luminosities \tabnotemark{a}} 
  \label{tab:1}  
  \begin{tabular}{lllcccccccc}
    \toprule
    &  &  &   \multicolumn{2}{c}{All} & & \multicolumn{2}{c}{Quasars} & & \multicolumn{2}{c}{BL Lac}\\
\cmidrule{4-5}\cmidrule{7-8}\cmidrule{10-11}

 A1 & A2 & A3   & $\tau$ & $P$ & & $\tau$ & $P$ &&  $\tau$ & $P$ \\

    \midrule
 $L_{5100}$	&	$ L_{\rm VLBA}$	&	z	&	\textbf{	0.259	}	&	\textbf{	3.95E-14	}	&	&	\textbf{	0.237	}	&	\textbf{	7.67E-10	}	&	&		0.140		&		0.173		\\
$L_{5100}$	&	$ L_{\rm un}$	&	z	&	\textbf{	0.242	}	&	\textbf{	6.14E-12	}	&	&	\textbf{	0.198	}	&	\textbf{	9.02E-07	}	&	&		0.150		&		0.149		\\
$L_{5100}$	&	$ L_{\rm jet}$	&	z	&	\textbf{	0.246	}	&	\textbf{	2.07E-12	}	&	&	\textbf{	0.218	}	&	\textbf{	6.53E-08	}	&	&	\textbf{	0.389	}	&	\textbf{	9E-04	}	\\

\bottomrule
   \tabnotetext{a}{The columns are arranged as follows: A1 and A2 are the independent variables for which the Kendall's $\tau$ correlation analysis is performed, and A3 is the dependent variable, in this case the redshift, ``All'' refers to the 233 sources in the sample, also were analyzed separately 181 quasars  and 31 BL Lacs, $\tau$ is the correlation coefficient, and $P$ is the probability of a chance correlation. The correlations considered to be significant are marked in bold face.}
  \end{tabular}
\end{table*}

\begin{figure*}[!t]
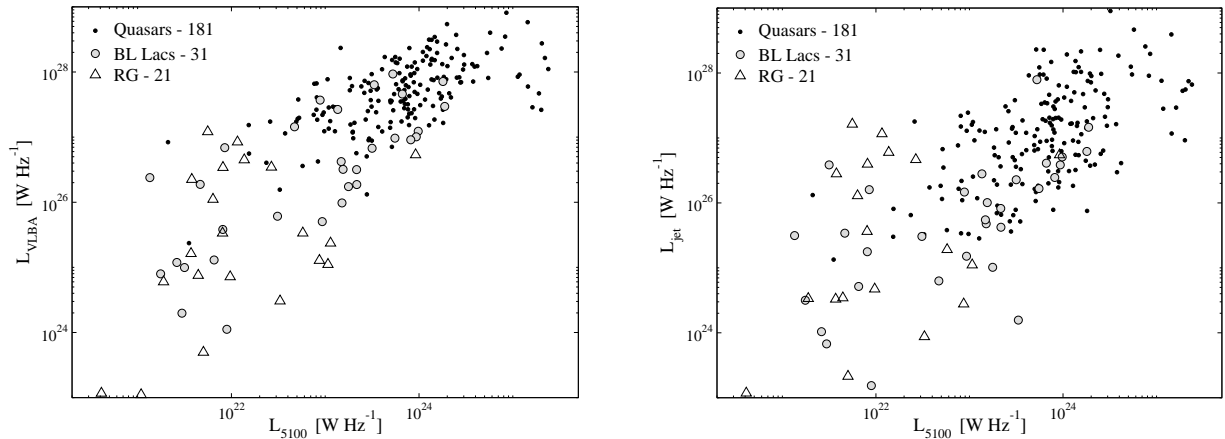

 
   \includegraphics[width=0.45\textwidth]{Lopt_LVLBA}%
    \hspace*{\columnsep}%
  \includegraphics[width=0.45\textwidth]{Lopt_Ljet}%
  \caption{\small{The total radio luminosity of the VLBA jet at 15\,GHz against optical continuum luminosity at 5100\,\AA\, (left panel), and radio luminosity of the jet (unresolved core subtracted) against optical continuum luminosity at 5100\,\AA\ (right panel) for quasars, BL Lacs and radio galaxies.}}
  \label{fig:1}
\end{figure*}

\section{Radio-Optical Correlations: analysis and summary}
\label{sec:results}
The total luminosity of the VLBA component ($L_{\rm VLBA}$), the unresolved core  ($L_{\rm un}$) and the jet luminosity ($L_{\rm jet}$) at 15 GHz were estimated as described in A10 \S~3. Optical nuclear luminosities at 5100 \AA\,($L_{5100}$) corrected for stellar contribution (Equation~(10) in A10) were estimated for 233 AGN of our sample from an homogeneous calibration using the B-band in the standard Johnson's photometric system. 

The partial Kendall's $\tau$ test \citep{akritas96} is used to check the correlation between the radio luminosities of the unresolved core and the jet and the optical nuclear luminosity. We consider the correlations to be significant for the samples of all 233 AGN and 181 quasars if the chance probability of the correlation $P<0.02$ (or confidence level $>98\,\%)$, and $P<0.05$ (or confidence level $>95\,\%)$ for the sample of 31 BL Lacs.  
The outcome of the statistical analysis is summarized in Table~\ref{tab:1} and they are in agreement with the results reported in A10. There is a significant positive correlation for 233 AGN between optical nuclear emission and radio emission  originated in the jet at milliarcseconds scales (Figure~\ref{fig:1}; Table~\ref{tab:1}). For quasars, correlations hold also between optical nuclear luminosity and radio luminosities of the unresolved core (at sub-milliarcseconds scales) and the whole jet. For BL Lacs, the optical luminosity correlates positively only with the jet luminosity at the high confidence level of $\sim\,99.9\%$ (Figure~\ref{fig:1}). 
The dispersion of the radio-optical correlation can be caused by non-simultaneous observations, distributions of intrinsic luminosities, and Doppler factors. The larger dispersion for BL Lacs could be a result of stronger variability in the radio and optical bands as well as a wider range of intrinsic luminosities, while for radio galaxies the deeming of optical continuum emission by an obscuring dusty torus can vary significantly.

Our results, together with the apparent speed -- optical luminosity diagram and the aspect curves derived from the relativistic beaming theory (see Figure~9 of A10), support the idea that the optical emission is generated in the relativistic jet. In particular, the optical emission in quasars with superluminal jets originates in the innermost part of the jet at sub-parsec scales, while in the BL Lacs it is generated in the parsec-scale jet.

\bibliographystyle{rmaa} 

\end{document}